\documentstyle[sprocl,epsf]{article}

\def\Journal#1#2#3#4{{#1} {\bf #2}, #3 (#4)}

\def\NPB{{\em Nucl.\ Phys.} B}
\def\PLB{{\em Phys.\ Lett.} B}

\def\PRD{{\em Phys.\ Rev.} D}
\def\ZPC{{\em Z.\ Phys.} C}
\def\RMP{{\em Rev.\ Mod.\ Phys.}}

\def\vslash{v\!\!\!\slash}

\begin{document}

\vbox{\rightline{CALT-68-2111, hep-ph/9704370}\hbox{}}

\title{NONPERTURBATIVE EFFECTS IN INCLUSIVE $\bar B\to X_s\,\gamma$
\footnote{Talk given at the FCNC'97 workshop, 
  February 19-21, 1997, Santa Monica.  }
}

\author{ZOLTAN LIGETI}

\address{California Institute of Technology, Pasadena, CA 91125}

\maketitle

\abstracts{Uncertainties in the theoretical prediction for the inclusive 
$\bar B\to X_s\,\gamma$ decay rate are examined.  Certain nonperturbative
effects involving a virtual $c\,\bar c$ loop, which are calculable using the
operator product expansion, are discussed.}

The inclusive $\bar B\to X_s\,\gamma$ decay is sensitive to physics beyond the
standard model, and the photon spectrum can help us better understand
nonperturbative effects in other $B$ decays.  As the CLEO data~\cite{CLEO}
excludes large deviations from the standard model, it is important to know the
theoretical predictions as precisely as possible.  With the completion of the
next-to-leading order calculation~\cite{Hurth}, the theoretical
uncertainty in perturbation theory is only about~10\%.

The effective weak interaction Hamiltonian at a scale $\mu(\sim m_b)$ is
$H_{\rm eff}=-(4G_F/\sqrt2)\,V_{ts}^*\,V_{tb}\,\sum_{i=1}^8C_i(\mu)\,O_i(\mu)$.
Here $O_2=(\bar s_{L\alpha}\,\gamma_\mu\,b_{L\beta})\, (\bar
c_{L\beta}\,\gamma^\mu\,c_{L\alpha})$, $O_1$ differs from $O_2$ in the
contraction of color indices, $O_3-O_6$ are other four-quark operators, and
$O_7=(e/16\pi^2)\,m_b\,\bar s_L\,\sigma^{\mu\nu} F_{\mu\nu}\,b_R$.

For large enough photon energies, the matrix element of $C_7\,O_7$ dominates
the $\bar B\to X_s\,\gamma$ rate.  Since $m_b\gg\Lambda_{\rm QCD}$, this
contribution is calculable by performing an operator product expansion (OPE)
for the time ordered product~\cite{FLS} 
\begin{equation}\label{ope1}
T_{77} = {i\over2m_B}\,\int {\rm d}^4x\, e^{-iq\cdot x}\, \langle\bar B(v)|\, 
  T\{O_7^{\mu\dagger}(x)\, O_7^\nu(0)\} |\bar B(v)\rangle\, g_{\mu\nu} \,.
\end{equation}
Here $O_7^\mu=(i\,e/8\pi^2)\,m_b\,\bar s_L\,\sigma^{\mu\lambda}q_\lambda\,b_R$.
At fixed $q^2=0$, $T_{77}$ has cuts in the complex $v\cdot q$ plane along
$v\cdot q<m_b/2$ and $v\cdot q>3m_b/2$ corresponding to final hadronic states
$X_s$ and $X_{bb\bar s}$, respectively.  The $\bar B\to X_s\,\gamma$ decay rate
is given by the discontinuity across the cut in the region $0<v\cdot q<m_b/2$,
\begin{equation}\label{77spec}
{{\rm d}\Gamma\over{\rm d}E_\gamma} = {4G_F^2\, |V_{ts}^*\,V_{tb}|^2\, C_7^2
  \over \pi^2}\, E_\gamma\, {\rm Im}\, T_{77}\,.
\end{equation}
Since the cuts are well-separated, this contribution can be computed assuming
local duality at the scale $m_b$ ($m_{X_s}=m_B$ at $v\cdot q=0$).  

At leading order in the OPE, the dimension-three operator $\bar
b\,\gamma_\mu\,b$ occurs.  Its matrix element gives a calculable contribution
proportional to $\delta(E_\gamma-m_b/2)$, which is equal to the free quark
decay result.  Higher dimension operators give terms proportional to
derivatives of this delta function.  The leading nonperturbative corrections
suppressed by $\Lambda_{\rm QCD}^2/\,m_b^2$ are quite small, about
$-3$\%~\cite{FLS}.  To justify retaining only the lowest dimension operators
whose matrix elements are known, the photon energy must be averaged over a
region $\Delta E_\gamma\gg\Lambda_{\rm QCD}$, and cannot be restricted to be
too close to its maximal (i.e., end-point) value.  Currently this is a source
of significant uncertainty, since the photon spectrum is only measured over a
region about $500\,$MeV from the end-point~\cite{CLEO}.

When operators in $H_{\rm eff}$ other than $O_7$ are included, there are
contributions from diagrams in which the photon couples to light quarks. 
Typically, the leading logarithms are calculable for such processes~\cite{ed},
but there are uncalculable contributions suppressed by a logarithm (or
equivalently by $\alpha_s$, but not by a power of the scale of the process),
which can only be estimated using the fragmentation functions $D_{q\to\gamma
X}$ and $D_{g\to\gamma X}$ deduced from other experiments or from models. 
Perturbative computations indicate that the contribution of light quark
loops~\cite{lql} and the effects related to decay functions of light partons
into a photon~\cite{KLP} are both very small for decays into hard
photons.~\footnote{For soft photons these are important.  Interference effects
where the photon couples to a light quark and either to the charm quark or
through $O_7$ are also small for hard photons.}  Assuming that these
calculations provide correct order of magnitude estimates~\cite{lore}, such not
power suppressed effects constitute less than five percent uncertainty in the
theoretical prediction for the $\bar B\to X_s\,\gamma$ decay rate.

Nonperturbative effects involving the photon coupling to the charm quark
contain matrix elements of local operators~\cite{Volo} only suppressed by
$\Lambda_{\rm QCD}^2/m_c^2$ rather than $\Lambda_{\rm QCD}^2/m_b^2$.  Such
effects could be sizable, since $m_b^2/m_c^2\sim10$ and the $\Lambda_{\rm
QCD}^2/m_b^2$ corrections are 3\%.  For a sufficiently heavy charm quark,
nonperturbative corrections to the interference of $O_2$ and $O_7$ can be
computed from the discontinuity of the diagram in Fig.~1.  Analogous diagrams
with more gluons are suppressed by additional powers of $\Lambda_{\rm
QCD}/m_c$.  Denote the photon and gluon momenta by $q$ and $k$ respectively. 
Working to all orders in $k\cdot q/m_c^2$ since $|q|\sim m_b$, but neglecting
$k\cdot q/m_b^2$, $k^2/m_{c,b}^2$, and $m_s/m_b$, gives~\cite{us}
\begin{equation}\label{mxel}
T_{27} = -{1\over2m_B}\, \langle\bar B(v)|\, \bar b\;
  m_b\,\sigma^{\nu\rho}q_\rho\,
  {m_b \vslash-q\!\!\!\slash\over(m_bv-q)^2+i\epsilon}\, 
  \gamma^\mu(1-\gamma_5)\, I_{\mu\nu}\, b\, |\bar B(v)\rangle \,.
\end{equation}
Here $I_{\mu\nu}$ is a complicated operator involving all powers of 
$q\cdot iD/m_c^2$,
\begin{equation}\label{Imunu}
I_{\mu\nu} = \bigg({e\over16\pi^2}\bigg)^2\, {2\over9m_c^2}\,
  \bigg[ \sum_{n=0}^\infty {3\,[(n+1)!]^2\,2^{n+3}\over(2n+4)!}\,
  \bigg({-q\cdot iD\over m_c^2}\bigg)^{\!n} \bigg]\,
  \varepsilon_{\mu\nu\lambda\beta}\, q^\beta q_\eta\, g_sG^{\lambda\eta} \,.
\end{equation} 
The covariant derivatives, $D$, act on the gluon field $G^{\lambda\eta}$, so
the matrix elements of these operators are determined by the spacetime
dependence of the chromomagnetic field in the $B$ meson.  The contribution of
$T_{27}$ to the $\bar B\to X_s\,\gamma$ decay rate is given by
Eq.~(\ref{77spec}) with $C_7^2\,T_{77}$ replaced by $2C_2C_7\,T_{27}$.  

\begin{figure}[t]  
\centerline{\epsfysize=2truecm \epsfbox{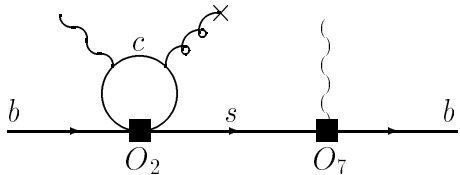}}
\caption[1]{Feynman diagram that gives rise to $T_{27}$ in Eq.~(\ref{mxel}).  
Interchange of the photon and gluon couplings to the charm loop is understood.}
\end{figure}

For the leading $n=0$ term in Eq.~(\ref{Imunu}), the matrix element in 
Eq.~(\ref{mxel}) is known from the $B^*-B$ mass splitting.  This gives
\begin{equation}
{\delta\Gamma(\bar B\to X_s\,\gamma) \over \Gamma(\bar B\to X_s\,\gamma)} = 
  -{C_2\over9C_7}\, {\lambda_2\over m_c^2} \simeq 2.5\% \,.
\end{equation}
Here we used $C_2=1.11$, $C_7=-0.32$, $\lambda_2=0.12\,{\rm GeV}^2$, and
$m_c=1.4\,$GeV.  This result is an order of magnitude larger than the
perturbative estimate of the $O_2\,O_7$ contribution (which contains a gluon in
the final state).  The $n=1$ matrix element vanishes due to the equations of
motion~\cite{GMNP}.  The $n>1$ terms in Eq.~(\ref{Imunu}) depend on an 
infinite series of unknown matrix elements.  Estimating 
$
\langle\bar B(v)|\, \bar b\, \Gamma^{\alpha\beta}(\widehat q,v)\, 
  (iD_{\mu_1}\ldots iD_{\mu_n}\, g_sG_{\alpha\beta})\, b\, 
  |\bar B(v)\rangle /(2m_B) \sim (\Lambda_{\rm QCD})^{n+2} 
$,
the $n>1$ terms are ``suppressed" compared to the $n=0$ term only by powers 
of $m_b\,\Lambda_{\rm QCD}/m_c^2$.  (A different estimate was given by Grant 
{\it et al}.~\cite{GMNP})

In the limit where $m_c$ is fixed and $m_b\to\infty$, the higher order terms in
Eq.~(\ref{Imunu}) become successively more important and the expansion we have
made is inappropriate.  (The whole contribution in Eq.~(\ref{mxel}) is still
suppressed by $\Lambda_{\rm QCD}/m_b$.)  In the limit where $m_b/m_c$ is fixed
and $m_{c,b}\to\infty$, the $n=0$ result dominates the sum in Eq.~(\ref{Imunu})
as the $n>1$ terms are suppressed by powers of $\Lambda_{\rm QCD}/m_c$.  In the
physical world, $m_b\,\Lambda_{\rm QCD}/m_c^2\sim0.6$ is of order unity.  As
the coefficients of the operators in Eq.~(\ref{Imunu}) are already small for
small $n$, and decrease asymptotically as $3\sqrt\pi/(2^{n+1}\,n^{3/2})$, the
$n>1$ terms probably do not introduce an uncertainty larger than the size of
the leading $n=0$ term.  Nonperturbative effects from the $O_1\,O_7$
interference are expected to be smaller.

Consider next the contribution of $(C_1O_1+C_2O_2)^2$.  Diagrams like that in
Fig.~1 should give a smaller result than for the interference of $O_2$ with
$O_7$ (of order $\Lambda_{\rm QCD}^4/m_c^4$ instead of $\Lambda_{\rm
QCD}^2/m_c^2$).  But in this case there is a contribution to the $\bar B\to
X_s\,\gamma$ decay rate from $\bar B\to X_s\,J/\psi$ followed by
$J/\psi\to\gamma\,X$, which is much larger than the perturbative calculation of
the effect of $(C_1O_1+C_2O_2)^2$.  The combined branching ratio for this
process is about as large as the total $\bar B\to X_s\,\gamma$ decay rate: of
order $10^{-4}$.  This might not present a serious difficulty for the
comparison of experiment with theory, since $\bar B\to X_s\,J/\psi$ followed by
$J/\psi\to\gamma\,X$ does not favor hard photons.  Moreover, it can be treated
as a background and subtracted away.  However, if such a subtraction is made,
is it not double counting to include the perturbative result for the
$(C_1O_1+C_2O_2)^2$ contribution into the theoretical prediction?  In any
event, $B\to X_s\,J/\psi$ followed by $J/\psi\to\gamma\,X$ is a long distance
contribution, while the charm quarks are far off-shell when nonperturbative
effects suppressed by powers of $\Lambda_{\rm QCD}/m_c$ are calculable. 
Further work on these issues is warranted.

In summary, the nonperturbative contribution to the matrix element of
$C_7\,O_7$, of order $\Lambda_{\rm QCD}^2/m_b^2$, is about $-3$\%~\cite{FLS}
with small uncertainty.  Although there is no OPE for the contribution of
photon coupling to light quarks, such effects give less than five percent
uncertainty for hard photons.  For the contribution of photon coupling to charm
quarks, there are nonperturbative effects of order $\Lambda_{\rm QCD}^2/m_c^2$,
whose magnitude is $2.5$\%, with a similar uncertainty.  However, it is
possible that larger nonperturbative effects may exist.  For comparison with
the present data~\cite{CLEO} which focuses on the region
$E_\gamma\geq2.2\,$GeV, the largest theoretical uncertainty is due to the
contribution of higher dimension operators to $T_{77}$ which become important
in the end-point region.  This uncertainty would be substantially smaller if
the photon energy cut were reduced.

I thank Lisa Randall and Mark Wise for collaboration.
This work was supported by the U.S.\ Dept.\ of Energy under 
grant no.\ DE-FG03-92-ER~40701.

\end{document}